\documentclass[letterpaper, 10 pt, conference]{ieeeconf}  

\IEEEoverridecommandlockouts                              

\usepackage[utf8]{inputenc}
\usepackage{soul,color}
\usepackage{amsmath,amssymb,amsfonts}
\usepackage{mathtools}
\usepackage{graphicx}
\usepackage{array}\newcolumntype{M}[1]{>{\centering\arraybackslash}m{#1}}
\usepackage{float} 
\usepackage{cite} 
\usepackage{subfigure}
\usepackage{tikz} 
\usepackage{enumerate}
\usepackage[algo2e,norelsize,linesnumbered,ruled]{algorithm2e}
\usepackage{balance} 
\usepackage{comment}
\usepackage{accents}
\usepackage{makecell}
\usepackage{multirow}
\usepackage{graphicx}
\newcommand{\ubar}[1]{\underaccent{\bar}{#1}}
\DeclareMathAlphabet{\mathcal}{OMS}{cmsy}{m}{n} 
\usepackage{flushend} 
\usepackage{textcomp, gensymb}
\usepackage{amsmath}

\title{\LARGE \bf
Switched Moving Boundary Modeling of Phase Change Thermal Energy Storage Systems*
}
\title{Switched Moving Boundary Modeling of Phase Change Thermal Energy Storage Systems}
\author{Trent J. Sakakini and Justin P. Koeln}
\date{January 2023}

\author{Trent J. Sakakini and Justin P. Koeln
\thanks{*This work was not supported by any organization}
\thanks{The authors are with the Department of Mechanical Engineering, The University of Texas at Dallas, 800 W. Campbell Rd, Richardson, TX, USA. Email: {\tt\small(trent.sakakini, justin.koeln)@utdallas.edu}.}}%

\begin{document}

\maketitle
\thispagestyle{empty}
\pagestyle{empty}

\begin{abstract}

Thermal Energy Storage (TES) devices, which leverage the constant-temperature thermal capacity of the latent heat of a Phase Change Material (PCM), provide benefits to a variety of thermal management systems by decoupling the absorption and rejection of thermal energy. While performing a role similar to a battery in an electrical system, it is critical to know when to charge (freeze) and discharge (melt) the TES to maximize the capabilities and efficiency of the overall system. Therefore, control-oriented models of TES are needed to predict the behavior of the TES and make informed control decisions. While existing modeling approaches divide the TES in to multiple sections using a Fixed Grid (FG) approach, this paper proposes a switched Moving Boundary (MB) model that captures the key dynamics of the TES with significantly fewer dynamic states. Specifically, a graph-based modeling approach is used to model the heat flow through the TES and a MB approach is used to model the time-varying liquid and solid regions of the TES. Additionally, a Finite State Machine (FSM) is used to switch between four different modes of operation based on the State-of-Charge (SOC) of the TES. Numerical simulations comparing the proposed approach with a more traditional FG approach show that the MB model is capable of accurately modeling the behavior of the FG model while using far fewer states, leading to five times faster simulations. 

\end{abstract}

\section{Introduction}
The need for higher performance and more efficient thermal management systems has driven the design of systems with integrated Thermal Energy Storage (TES) devices that leverage the latent heat of a Phase Change Material (PCM). The design and performance of PCM-based TES has been well-studied \cite{Nazir2019,Sharma2009,Zalba2003}, resulting in a wide range of applications including building \cite{Lee2020,Ma2012} and aircraft \cite{Laird2021,Laird2019} thermal management, power electronics cooling \cite{Pangborn2020}, and combined heating and cooling \cite{Bird2020}.

The utility of a TES is heavily dependent on the dynamics associated with charging (where the PCM solidifies from liquid to solid), discharging (where the PCM melts from solid to liquid), and strategic switching between these two modes of operation. Therefore, accurate control-oriented models of PCM-based TES are needed that capture their hybrid, nonlinear dynamics to be used in predictive controllers like Model Predictive Control (MPC), which have been developed for single-phase \cite{Ma2012,Lee2020} and phase change TES \cite{Pangborn2020}. 


Traditional TES modeling approaches rely on dividing the PCM into multiple sections, where each section is modeled using a lumped-parameter approach.  This Fixed Grid (FG) approach, also referred to as Finite Volume, is widely used in the literature \cite{Shanks2022,Pangborn2015,Fasl2013} and is similar to a finite difference scheme \cite{Fortunato2012}. While this approach has proven to accurately model the complex dynamics of a TES device using relatively simple dynamics for each individual grid section, a large number of grid sections is needed to achieve this accuracy, resulting in a large number of dynamic states that is no longer practical for many control designs.


This paper aims to develop accurate control-oriented models of PCM-based TES devices using a graph-based switched Moving Boundary (MB) approach. Graph-based modeling \cite{Wang2020,Laird2021,Laird2019,Pangborn2020,Shanks2022} is used to develop both FG and MB models, where a graph is used to clearly identify the underlying structure of thermal energy storage and transfer throughout the TES device.  While the FG model divides the PCM into $ n $ sections, each with its own dynamic enthalpy state, the proposed MB approach only requires three states corresponding to the enthalpies of the solid and liquid regions of the PCM and the overall State-of-Charge (SOC), defined as the mass of the solid portion compared to the total mass of the PCM.

\begin{figure*}[t]
    \centering
    \includegraphics[width=\textwidth] {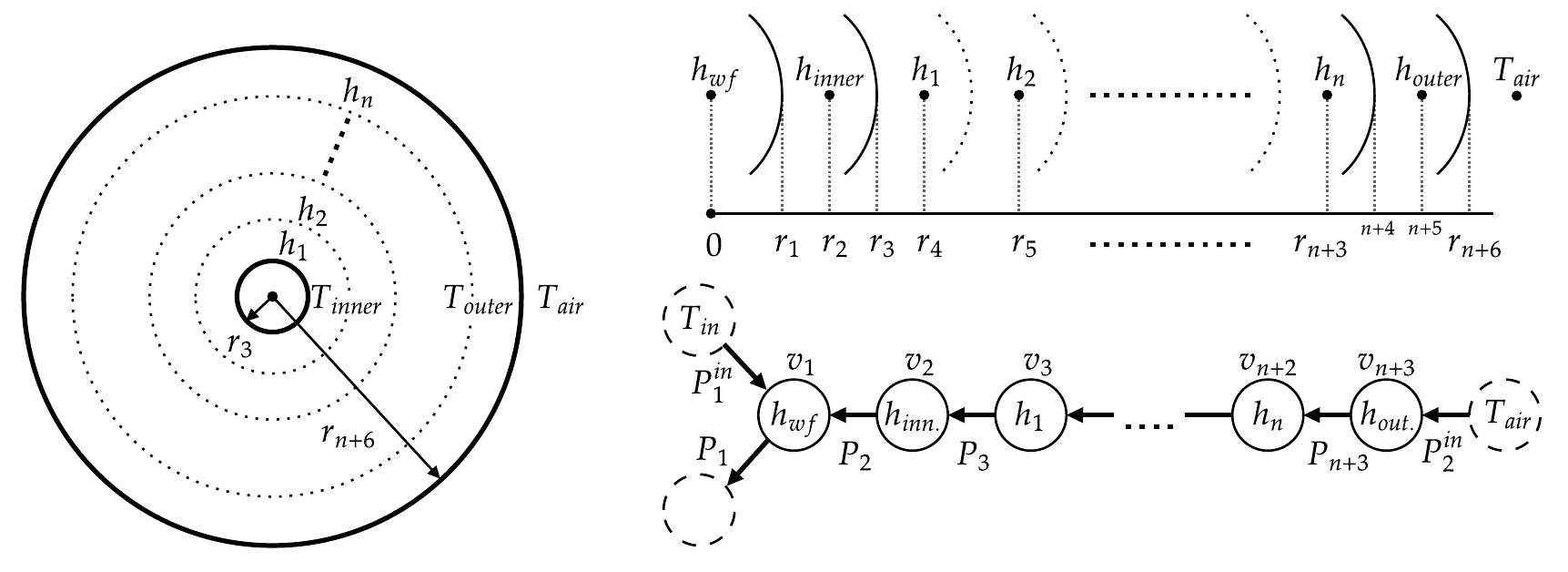}
    \vspace{-22pt}
    \caption{Fixed Grid modeling framework. 
    \textbf{LEFT:} Cylindrical TES with inner and outer walls and the PCM divided into $n$ grid sections. 
    \textbf{TOP RIGHT:} Identification of key radii used to model the 1-dimensional radial heat transfer.
    \textbf{BOTTOM RIGHT:} Graph-based FG model with $ n $ PCM vertices.}
    \label{fig:Fixed_Grid_Modeling}
    \end{figure*}
    
Several MB approaches to TES modeling have recently been proposed \cite{Fasl2013,Laird2019,Pangborn2020} but each has limitations. Specifically, the TES devices modeled in both \cite{Laird2019} and \cite{Fasl2013} are limited to operation where heat flows in only one direction through the PCM, i.e., heat always enters on one side and exits on the other. However, many TES devices operate by exchanging heat with a single working fluid flowing through the center of the TES. The TES model developed in \cite{Pangborn2020} captures how a heat transfer effectiveness coefficient for the PCM is a function of SOC and the mode of operation (charging or discharging) but modeling the completely melted or solidified modes of operation is left for future work.



The proposed switched MB approach overcomes these limitations using a Finite State Machine (FSM) to model the mode-dependent dynamics associated with freezing, melting, completely solid, and completely liquid operation. A similar FSM is used to model the operation of ultracapacitors \cite{Dede2016}. When compared to the FG model in a simulated example, the proposed switched MB model achieves a maximum error of $5\%$ for the SOC of the PCM with a $80\%$ reduction in computational time. Therefore, the primary contribution of this paper is the specific formulation of a graph-based switched MB model that is able to accurately predicted the key dynamics of PCM-based TES devices with far fewer states when compared to a more traditional FG model.

The remainder of the paper is organized as follows. Section II introduces and motivates the graph-based modeling framework. The traditional FG approach for TES modeling is presented in Section III while the proposed switched MB approach is presented in Section IV. The simulation accuracy and computational efficiency of the proposed modeling framework is compared to the FG approach in Section V. Finally, conclusions and future work are summarized in Section VI.

\section{Graph-Based Modeling Framework}

This paper employs graph-based modeling to capture the storage and transfer of energy in PCM-based TES devices. Specifically, Fig. \ref{fig:Fixed_Grid_Modeling} shows the graph-based model using the FG approach for a cylindrical TES device.
This TES device consists of two concentric cylindrical pipes, where the PCM is encapsulated between the inner and outer pipes. A working fluid flows through the inner pipe with an inlet temperature $T_{in}$ and is the main mechanism in which heat is transferred between the TES device and the remainder of the overall thermal management system (which is not modeled in this work). The outer wall of the outer pipe is assumed to exchange heat with ambient air at temperature $ T_{air} $. In the FG approach, the PCM is divided to $ n $ grid sections, where the $i^{th}$ grid section is assumed to have a uniform enthalpy $ h_i $.  This paper only considers the radial heat transfer of the TES device and assumes uniform behavior along the length of the  device $ L $. While the proposed approach is intended to extend to TES of different geometries, only the one-dimensional radial case is considered in this paper.

When capturing the structured dynamics of a system, a graph consists of a set of $ N_v $ dynamic vertices $V = \{v_{i}:i\in[1,N_{v}]\}$, representing energy stored by capacitative sections of a system, and a set of $ N_e $ edges $E=\{e_{j}:j\in[1,N_{e}]\}$, representing power flows among these capacitative sections. Note that $ [1,N_{v}] $ is used to denote the set of integers between $ 1 $ and $ N_v $. Each edge $ e_j $ has an orientation denoting the direction of positive power flow $ P_j $ from the tail vertex $ v_{j}^{tail} $ to the head vertex $ v_{j}^{head} $. Based on conservation of energy, the energy stored by $i^{th}$ vertex $v_{i}$ (quantified by the dynamic state $x_{i}$) can be expressed as
\begin{equation} \label{eq:conservationOFEnergy}
	C_{i}{\dot x}_{i}=\sum\limits_{e_{j} \in {E_{i}}^{in}} P_{j}-\sum\limits_{e_{j} \in {E_{i}}^{out}} P_{j},
\end{equation}
where $ C_i $ is the energy storage capacitance while ${E_{i}}^{in}$ and ${E_{i}}^{out}$ are the set of edges directed into and out of vertex $v_{i}$. Generally, in a graph-based modeling framework, the power flow $P_{j}$ is constrained to be a function of an associated input $\tilde{u}_{j}$ and the state of the tail and head vertices, $x_{j}^{tail}$ and $x_{j}^{head}$, such that 
\begin{equation} \label{eq:powerFlow}
	P_{j}= f_{j}(x_{j}^{tail},x_{j}^{head},\tilde{u}_{j}).
\end{equation}

In general, the graph-based modeling framework allows for power to enter the system along source edges as discussed in \cite{Wang2020}.  For the TES device shown in Fig. \ref{fig:Fixed_Grid_Modeling}, there are two sources into the system: heat transfer with the main working fluid ($wf$) and with the surrounding air. 
For heat that is being transferred out of the system, a sink vertex denoted $V^{out} = \{v_{i}^{out}:i\in[1,N_{v}^{out}]\}$ is included in the graph. This vertex has an associated state $ x_i^{out} $ that serves as the outlet of the working fluid.

The structure of the graph, including both the dynamic vertices and sink vertices, is captured by the incidence matrix $ M = [m_{ij}] \in \mathbb{R}^{(N_v + N_v^{out}) \times N_e } $ defined as
\begin{equation} \label{eq:Incidence_Matrix_Creation}
m_{ij}=  \begin{cases}
+1 & \text{if } v_i \text{ is the tail of } e_j, \\
-1 & \text{if } v_i \text{ is the head of } e_j, \\
0 & \text{else}.
\end{cases}
\end{equation}
The incidence matrix is partitioned based on dynamic and sink vertices such that
\begin{equation}
	M = \begin{bmatrix} \bar{M} \\ \ubar{M} \end{bmatrix} \text{ with } \bar{M} \in \mathbb{R}^{N_v \times N_e },
\end{equation}
where the indexing of vertices is assumed to be ordered such that $ \bar{M} $ is a structural mapping from power flows 
\begin{equation} \label{eq:powerFlows}
P = F(x,x^{out},\tilde{u}) = [f_j(x_j^{tail},x_j^{head},\tilde{u}_j)],
\end{equation}
to states $ x = [x_i] $, $ i \in [1,N_v] $, and $ \ubar{M} $ is a structural mapping from $ P $ to sink states $ x^{out} = [x_i^{out}] $, $ i \in [1,N_v^{out}] $. 
Combining the individual conservation equations from \eqref{eq:conservationOFEnergy} using the structure of the graph captured by $ \bar{M} $, the overall system dynamics are
\begin{equation} \label{eq:graphDynamics}
C \dot{x} = -\bar{M} P =  -\bar{M} F(x,x^{out},\tilde{u}),
\end{equation}
where $ C = diag([C_i]) $, $ i \in [1,N_v] $ is a diagonal matrix of capacitances. Since some edges do not have a control input and a single input can affect multiple edges, it is often advantageous to let $ \tilde{u} \in \mathbb{R}^{N_e} $ be a virtual input vector, corresponding to the $ N_e $ edges, and define $ u \in \mathbb{R}^{N_u} $ as a system input vector, corresponding to the subset of $ N_u $ unique inputs that affect the system.  As such, the matrix $ \Phi \in \mathbb{R}^{N_e \times N_u} $ can be used to map the system inputs to the virtual inputs such that $ \tilde{u} = \Phi u $.

One benefit of a graph-based modeling framework is that the linear structure of the graph is captured by \eqref{eq:graphDynamics} and the majority of the modeling effort focuses on defining the potentially nonlinear power flow relationships in \eqref{eq:powerFlows}. The following section presents the graph capturing the structure of the system shown in Fig. \ref{fig:Fixed_Grid_Modeling} and the vertex and edge properties used to model the dynamics. 

\section{Fixed Grid TES Modeling Framework}
\subsection{Modeling Assumptions}
The dynamics of the TES, comprised of the working fluid, inner wall, PCM, and outer wall, are modeled using a graph-based framework with the following assumptions.
\begin{itemize}
    \item Heat transfer within the TES is radially symmetric and uniform along the length of the device.
    \item Heat transfer is assumed to be purely conductive. Natural convection in the liquid is not taken into account, similar to \cite{Laird2019}. Future experimental work similar to \cite{jalil2006,Kahraman1998} will focus on quantifying and incorporating the effects of natural convection into the graph-based modeling framework.
    \item The mass in the PCM is assumed to be constant with time-varying volume based on the density changes associated with phase change.
    \item Heat transfer between the working fluid and the inner pipe is governed by the outlet temperature of the working fluid.
    \item All material properties are phase dependent but constant within each phase.
    \item The pressure of the PCM is assumed to be constant over time, space, and phase and does not influence the TES dynamics.
\end{itemize}

The following graph-based models use enthalpies as system states, since temperature cannot be used to quantify thermal energy during phase change. The PCM is generically assumed to have a saturated solid state enthalpy of $ h = 0 \, kJ/kg $, a latent heat of fusion of $ h_f $, and a saturated temperature of $ T_{sat} $.
Temperature $T $ for the PCM is defined as
\begin{equation}
    T =  \begin{cases}
    \frac{h}{C_{p,\sigma}} + T_{sat} & \text{if } h < 0, \\
    T_{sat} & \text{if } 0 \leq h \leq h_f, \\
    \frac{h-h_f}{C_{p,\sigma}} + T_{sat} & \text{if } h > h_f, \\
    \end{cases}
    \end{equation}
where $C_{p,\sigma}$ is the phase-dependent specific heat capacity of the PCM and the phase
$\sigma$, either solid (S) or liquid (L), is 
\begin{equation}
    \sigma =  \begin{cases}
    S & \text{if } T < T_{sat}, \\
    L & \text{if } T \geq T_{sat}.
    \end{cases}
\end{equation}
For single-phase materials, such as the working fluid and the pipe walls, temperature is defined as $T = \frac{h}{C_p}$,
where $C_p$ is the specific heat capacity of the material.

\subsection{Fixed Grid Approach}
The traditional FG approach to modeling PCM-based TES devices divides the volume into $n$ grid sections \cite{Shanks2022,Fasl2013}. The FG modeling framework is used as a reference in this paper, representing the true dynamic behavior of the TES to be approximated by the proposed switched MB approach. 
As shown in Fig. \ref{fig:Fixed_Grid_Modeling}, the FG approach requires a total of $ n+3 $ states such that $ x \in \mathbb{R}^{n+3} $, where $ x = [h_{wf}, \, h_{inn.}, \, h_1, \, \dots, \, h_n, h_{out.} ]^\top $ are the enthalpies of the working fluid, the inner wall, the $ n $ sections of PCM, and the outer wall.


The following graph-based FG model is derived from the approach presented in \cite{Fasl2013} and the radial heat transfer equations from \cite{Bergman2011}. Modeling each vertex in Fig. \ref{fig:Fixed_Grid_Modeling} using conservation of energy, with state $ h_i $ for the $ i^{th} $ vertex, the energy storage capacitance $ C_i $ from \eqref{eq:conservationOFEnergy} is the mass of the vertex such that $ C_i = \rho_i  V_i $ for the single-phase material vertices $ i \in \{1, 2, n+3\} $ and $ C_i = \rho_{i,\sigma} V_i $ for the PCM vertices $ i \in \{3,n+2\} $. The density $ \rho_i $ is assumed constant for single-phase materials while $ \rho_{i,\sigma} $ denotes the fact that the PCM density is phase-dependent. The volumes $ V_i $ for the three single-phase vertices are defined as $ V_1 = \pi L r_1^2 $, $ V_2 = \pi L (r_3^2 - r_1^2) $, and $ V_{n+3} = \pi L (r_{n+6}^2 - r_{n+4}^2) $, based on the radii labelled in Fig. \ref{fig:Fixed_Grid_Modeling}, where $ L $ is the length of the TES device. The PCM is divided into $ n $ sections of equal width $ \Delta r = \frac{r_{n+4} - r_3}{n} $ such that the volumes $ V_i, \, i \in [3,n+2] $, are defined as $ V_i = \pi L [(r_{i+1} + \frac{\Delta r}{2})^2 - (r_{i+1} - \frac{\Delta r}{2})^2 ]$.

Each power flow $ P_j $ can be expressed in the form of \eqref{eq:powerFlow} assuming positive power flow in the direction of the arrows shown in Fig. \ref{fig:Fixed_Grid_Modeling}. The advective power flows associated with the working fluid are $ P^{in}_1 = \dot{m}_{wf} C_{p,wf} T_{in} $ and $ P_1 = \dot{m}_{wf} C_{p,wf} T_1 $,
where $\dot{m}_{wf}$ is the mass flow rate and $C_{p,wf}$ is the specific heat capacity of the working fluid. For heat transfer from the surrounding air into the TES, the outer wall is a combination of the pipe material and insulation,
\begin{equation}
     \begin{gathered}
     P^{in}_2 = \frac{1}{R_{out.} + R_{air}} (T_{air} - T_{n+3}),\\
     R_{out.} = \frac{ln(\frac{r_{n+6}}{r_{n+5}})}{2 \pi L k_{out.}} + R_{ins.},
     R_{air} = \frac{1}{2 \pi r_{n+6} L h_{air}},
    \end{gathered}
\end{equation}
where $ k_{out.} $ is the thermal conductivity of the outer pipe, $ h_{air} $ is the convective heat transfer coefficient for the air, and $R_{ins}$ is the insulation resistance. For power flows $ P_j, \, j \in [2,n+3] $,
\begin{equation} \label{eq:Power_Flow}
    P_j = \frac{1}{R_j} (T_{j} - T_{j-1}).
\end{equation}
Since each power flow $ P_j, \, j \in [2,n+3] $, goes through two different materials, the total thermal resistance is defined as $ R_j = R_{j,A} + R_{j,B} $, where $ R_{2,A} = \frac{1}{2 \pi r_{1} L h_{wf}} $, $ R_{2,B} = \frac{ln(\frac{r_2}{r_1})}{2 \pi L k_{inn.}}$, $ R_{3,A} = \frac{ln(\frac{r_3}{r_2})}{2 \pi L k_{inn.}}$, $ R_{n+3,B} = \frac{ln(\frac{r_{n+5}}{r_{n+4}})}{2 \pi L k_{out.}}$, and, $ \forall j \in [3,n+2] $,
\begin{equation}
    R_{j,B} = \frac{ln(\frac{r_{j+1}}{r_{j+1}-\frac{\Delta r}{2}})}{2 \pi L k_{j,\sigma}}, \; R_{j+1,A} = \frac{ln(\frac{r_{j+1}+\frac{\Delta r}{2}}{r_{j+1}})}{2 \pi L k_{j,\sigma}},
\end{equation}
where $ k_{inn.} $ is the thermal conductivity of the inner pipe and $ h_{wf} $ is the convective heat transfer coefficient for the working fluid.

\begin{figure}[t]
        \includegraphics[width=\columnwidth] {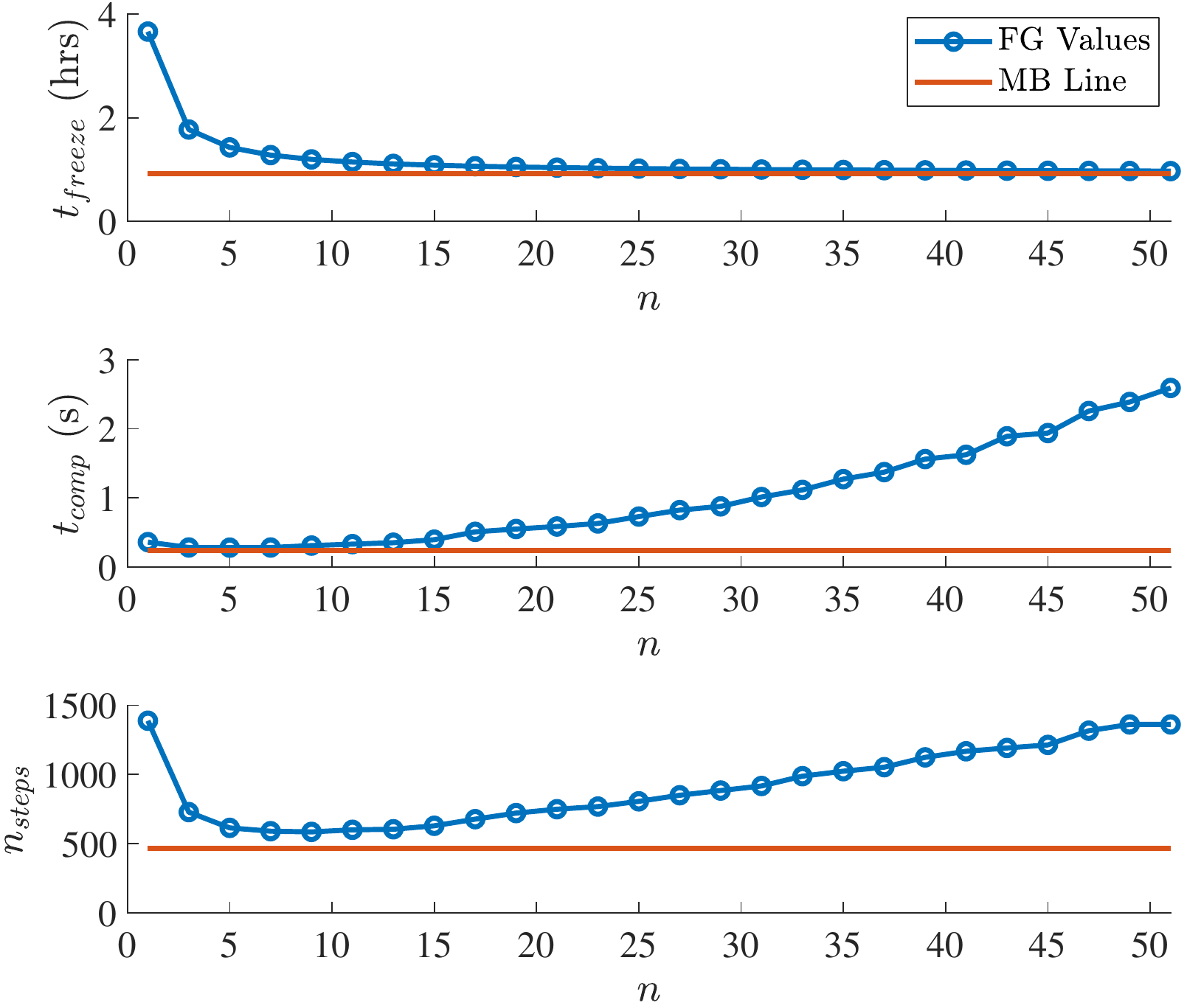}
        \vspace{-15pt}
        \caption{Computational comparisons of the FG and MB approaches. 
        \textbf{TOP:} Time the model estimates for the PCM to completely freeze, $t_{freeze}$. 
        \textbf{MIDDLE:} Computational time, $t_{comp}$. 
        \textbf{BOTTOM:} Number of time steps taken with the ode23tb variable step solver, $n_{steps}$. 
        All results are taken as an average over 50 simulations.}
        \label{fig:Freeze_Comp_Steps.pdf}
        \end{figure}

For the numerical example presented in Section \ref{Results}, Fig. \ref{fig:Freeze_Comp_Steps.pdf} shows the results of a series of tests to determine the behavior of the FG model as a function of $ n $. The top plot shows the simulated time required to completely freeze the TES, $t_{freeze}$, for different values of $ n $. While $t_{freeze}$ converges for increasing $ n $, the second plot shows the associated increase in computation time for the simulation, $t_{comp}$. These simulations were conducted in MATLAB Simulink using the variable step solver ode23tb. The third plot shows that the increase in computation time is due to an increasing number of states and simulation time steps. Based on the results of Fig. \ref{fig:Freeze_Comp_Steps.pdf}, $ n = 35 $ sections was chosen for comparison with the proposed switched MB approach presented in the following section.


\begin{figure*}[h]
    \centering
    \includegraphics[width=\textwidth] {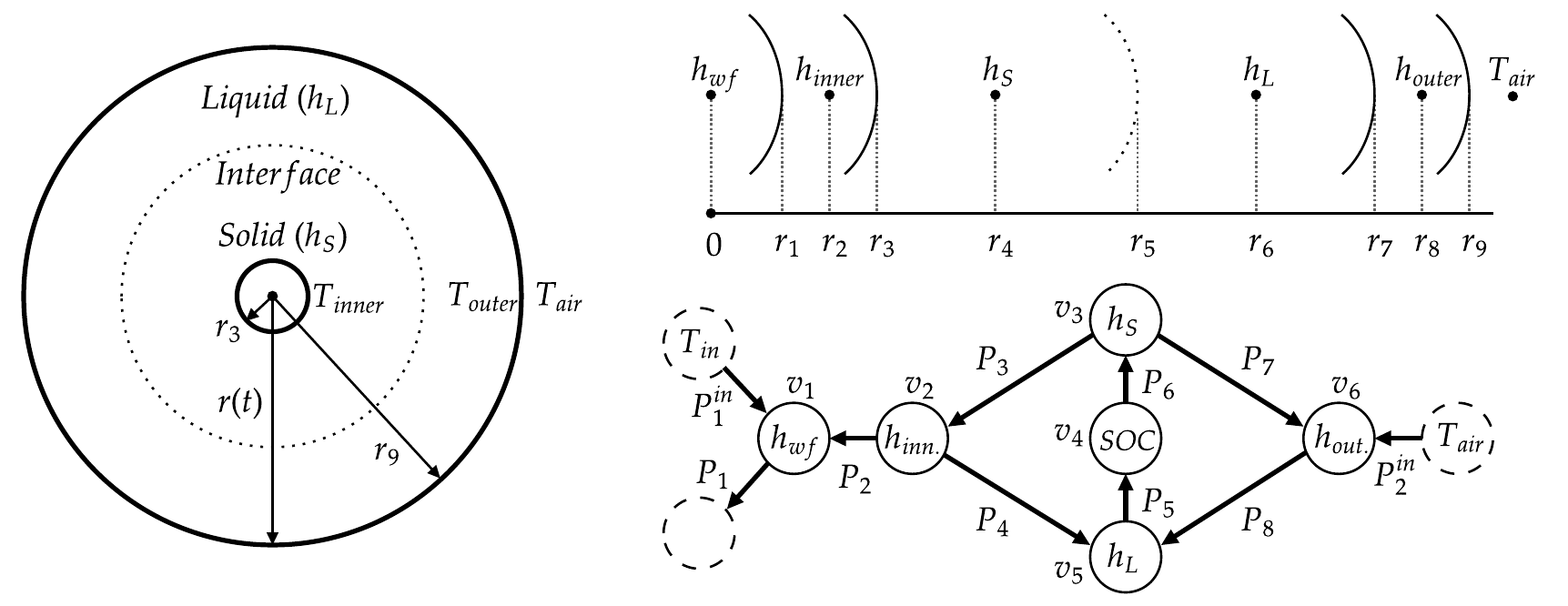}
    \vspace{-22pt}
    \caption{Proposed MB modeling framework. 
    \textbf{LEFT:} Cylindrical TES with inner and outer walls and the PCM divided into solid and liquid regions, with states $h_S$ and $h_L$, respectively. 
    \textbf{TOP RIGHT:} Identification of key radii used to model the 1-dimensional radial heat transfer.
    \textbf{BOTTOM RIGHT:} Graph-based MB model with three vertices for the PCM.}
    \label{fig:Moving_Boundary_Modeling}
    \end{figure*}

\section{Proposed Switched Moving Boundary Modeling Framework}
\subsection{Moving Boundary Model}
The MB approach aims to capture the primary dynamics of the TES using a reduced number of states to generate a model that can be directly used for control design.
As shown in Fig.~\ref{fig:Moving_Boundary_Modeling}, the proposed MB approach requires a total of six states such that $ x \in \mathbb{R}^{6} $, where $ x = [h_{wf}, \, h_{inn.}, \, h_S, \, SOC, \, h_L, \, h_{out.} ]^\top $ has only three PCM states corresponding to the enthalpies of the solid ($h_S$) and liquid ($h_L$) regions of the PCM as well as the SOC.
This can be significantly fewer states than the FG approach which requires $ n+3 $ states, where $ n = 35 $ was determined to be a practical balance between model accuracy and computational cost. The results of the MB mode are also presented in Fig. \ref{fig:Freeze_Comp_Steps.pdf} as the horizontal red lines, which show that the MB model accurately predicts $t_{freeze}$ with significantly less computation time and simulation steps.

Since the MB model only changes the configuration of the PCM, power flows $ P_j, \, j \in \{1,2\} $, power inputs $ P^{in}_j, \, j \in \{1,2\} $, capacitances $ C_i, \, i \in \{1,2,6\} $, and resistances $ R_{j,A}, \, j \in \{1,2,3\} $ and $ R_{j,B}, \, j \in \{1,2\} $ are all the same for the working fluid, and the inner and outer wall as defined in the FG model, by replacing $v_{n+3}$ in the FG with $v_6$ in the MB. 

 Modeling the three new vertices in Fig. \ref{fig:Moving_Boundary_Modeling} using conservation of energy, the energy storage capacitance $ C_i $ from \eqref{eq:conservationOFEnergy} is the mass of the vertex such that $ C_3 = M_{tot} SOC $, $C_4 = M_{tot} (h_S-h_L)$, and $C_5 = M_{tot} (1-SOC)$, where $M_{tot}$ is the total mass of the PCM, $SOC = \frac{M_S}{M_{tot}}$, and $M_S$ is the mass of the solid PCM. 
 While capacitances are typically positive in a graph-based modeling framework, $C_4 < 0 $ since $h_S < h_L $, which comes directly from the fact that the state, $ x_4 = SOC $, increases with a decrease in energy stored in the PCM such that $ SOC = 0 $ and $ SOC = 1 $ correspond to the PCM being completely liquid and completely solid, respectively. 
 
The power flow $ P_j, \, j \in [3,8] $, are defined similarly to \eqref{eq:Power_Flow}, such that power flow is driven by the temperature difference between the tail and head vertex temperatures for each edge. Note that $ T_{sat} $ is used as the vertex temperature for $ v_4 $ with state corresponding to SOC. The total thermal resistance is also still defined as $ R_j = R_{j,A} + R_{j,B} $ but now the radii associated with the solid and liquid regions are time varying. For example, as shown in Fig. \ref{fig:Moving_Boundary_Modeling}, $r_5 = \sqrt{r_3^2 + \frac{M_S}{\rho_S \pi L}}$.

The following section shows how a FSM is used to turn on and off power flows in Fig. \ref{fig:Moving_Boundary_Modeling} to accurately model the dynamics of the TES device under four distinct modes of operation.



\begin{figure}[t]
    \centering
    \includegraphics[width=\columnwidth] {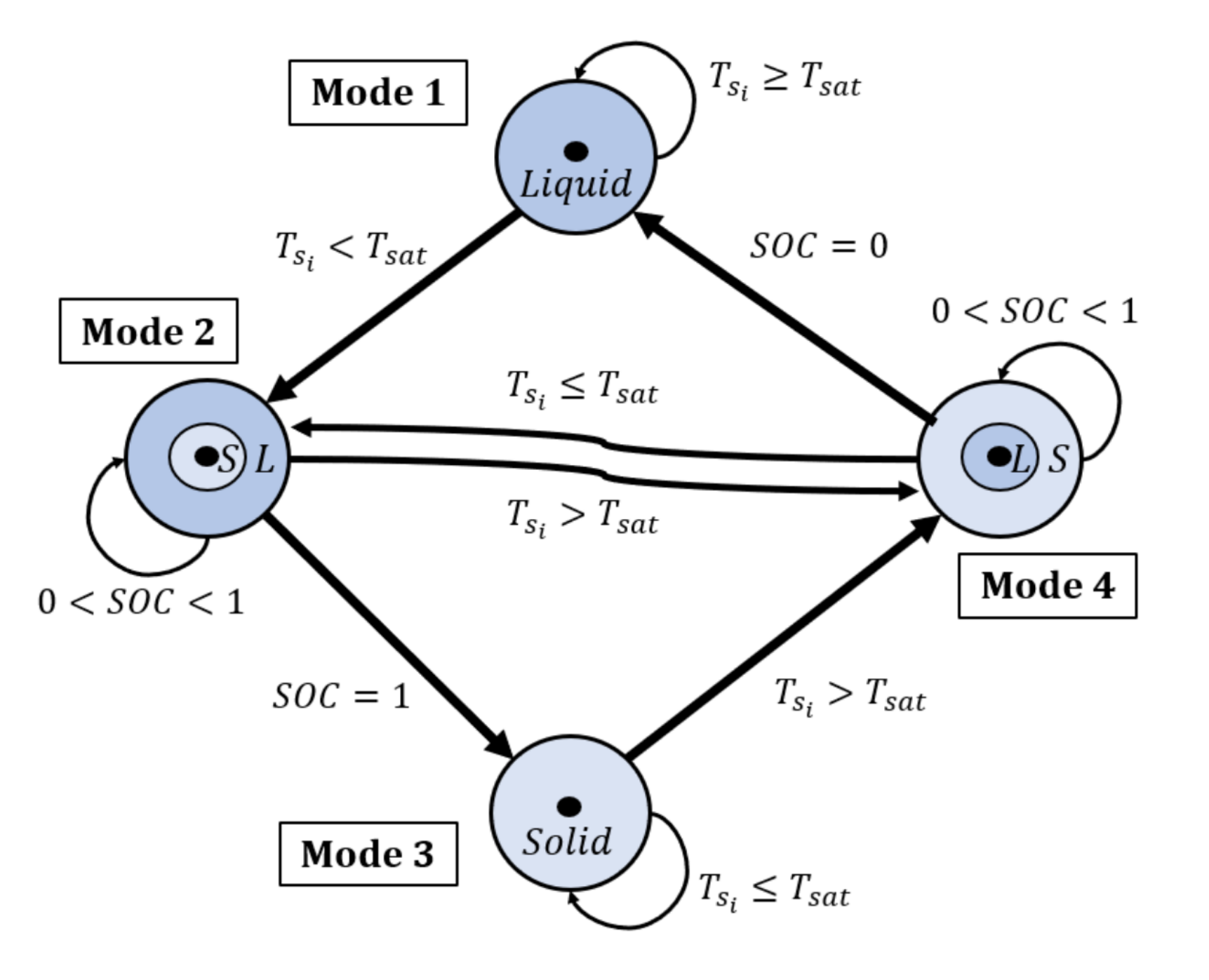}
    \vspace{-30pt}
    \caption{FSM with switching criteria for the four modes of the MB model.}
    \label{fig:Finite_State_Machine}
\end{figure}
    
\subsection{Finite State Machine}
The TES device has four major modes of operation: completely liquid, completely solid, freezing, and melting modes, as shown in Fig. \ref{fig:Finite_State_Machine} with their respective mode numbers. Mode switching is based on the SOC and the surface temperature $T_{s_i}$ between the inner pipe wall and the PCM defined as  
\begin{equation} \label{eq:Surface_Temperature}
    T_{s_i} =  \begin{cases}
    T_{inn.} + R_{3,A} P_3& \text{if } Mode \in \{2,3\}, \\
    T_{inn.} + R_{4,A} P_4& \text{if } Mode \in \{1,4\}.
    \end{cases}
    \end{equation}
    
Assuming the PCM starts in a completely liquid state (Mode~1), when the inlet working fluid temperature $ T_{in} < T_{sat} $ eventually $T_{s_i} < T_{sat}$ and the freezing process begins, switching the model into Mode 2 of the FSM. During the freezing process, the SOC will increase until the PCM is completely solid where $SOC = 1$ and the model switches to Mode 3. If the inlet working fluid temperature increases such that $ T_{in} > T_{sat} $, then eventually $T_{s_i} > T_{sat}$, and the melting process begins by switching to Mode 4. Once the PCM is complete liquid where $ SOC = 0 $, the model switches back into Mode 1. If the inlet working fluid temperature changes when the system is in Modes 2 or 4 before completely freezing or melting, the model can switch directly between Modes 2 and 4 with the PCM in a partially frozen state. During such transitions, note that the model makes a non-physical assumption that locations of the solid and liquid regions instantaneously switch such that solid is surrounded by liquid in Mode 2 and vice versa in Mode 4, as shown in Fig. \ref{fig:Finite_State_Machine}. While the SOC state still evolves continuously, the radii associated with the solid and liquid regions will change instantaneously.

While the graph in Fig. \ref{fig:Moving_Boundary_Modeling} shows all of the potential power flows through the PCM, power flows $ P_3 $ through $ P_8 $ are turned on and off based on the mode of operation as summarized in Table \ref{tab:Power_Flow_Modes}. For example, when the PCM is completely liquid (Mode 1), power flows $P_3$, $P_5$, $P_6$, and $P_7$ are all turned off to completely disconnect vertices $ v_3 $ and $ v_4 $ and allow both the inner and outer walls to exchange heat with only the liquid, vertex $ v_5 $. 

\begin{table}[t]
    \scriptsize
    \caption{\uppercase{Power Flows for each FSM Mode}}
    \centering
    \label{tab:Power_Flow_Modes}
    \begin{tabular}{p{1.5cm}<{\centering} p{0.8cm}<{\centering} p{0.8cm}<{\centering} p{0.8cm}<{\centering} p{0.8cm}<{\centering}}
     \hline
     Power Flow & Mode 1 & Mode 2 & Mode 3 & Mode 4 \\ [0.5ex] 
     \hline
     $P_3$ & off & on & on & off \\ 
     $P_4$ & on & off & off & on \\ 
     $P_5$ & off & on & off & on \\ 
     $P_6$ & off & on & off & on \\ 
     $P_7$ & off & off & on & on \\ 
     $P_8$ & on & on & off & off \\ [1ex] 
     \hline
    \end{tabular}
    \end{table}
Finally, the radii labeled in Fig. \ref{fig:Moving_Boundary_Modeling} only correspond to Mode 2 of the FSM and are used in computing the thermal resistances $R_{j,A}$ and $R_{j,B}$. For the other three modes, the equations for these thermal resistances must be modified to reflect the geometry and corresponding radii for each mode.  

    \begin{figure*}[t]
            \centering
            \includegraphics[width=\textwidth]{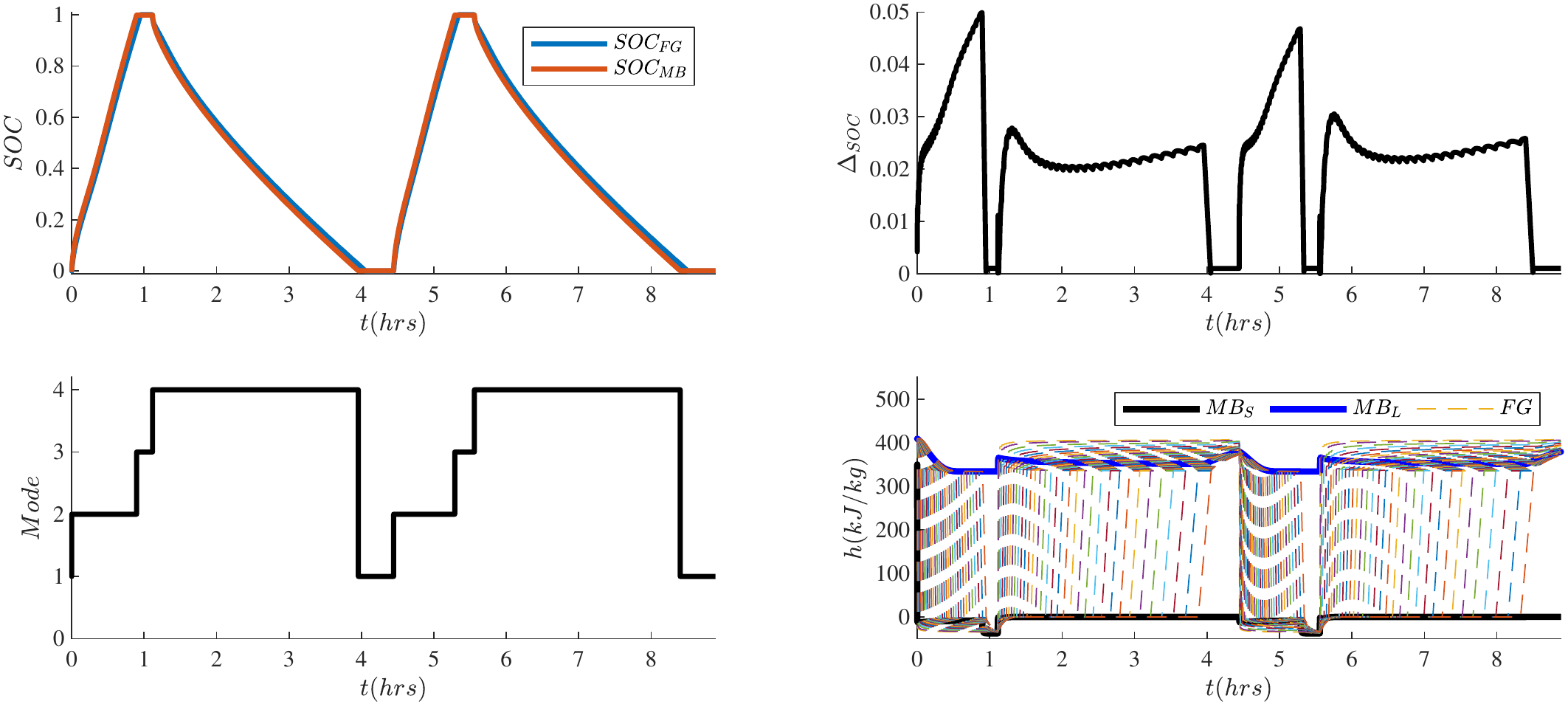}
            \caption{Differences between FG (with $n = 35$) and MB models for two complete freezing and melting cycles.}
            \label{fig:Freeze_Melt_Plot.pdf}
            \end{figure*}    

\section{Results} \label{Results}
\subsection{Simulation Setup}
Table \ref{tab:SimParams} shows the simulation parameters used to compare the proposed MB model with the more traditional FG model. The material properties for the simulated TES device assume that the working fluid is a 50/50 water-glycol mixture ($wg$), the inside pipe is copper ($Cu$), the PCM is water, and the outer pipe ($PVC$) is assumed to be well insulated.

\begin{table}[t]
    \scriptsize
    \caption{\uppercase{Simulation Parameters}}
    \centering
    \label{tab:SimParams}
    \begin{tabular}{p{0.9cm}<{\centering} p{3.9cm}<{\centering} p{1cm}<{\centering} p{1.1cm}<{\centering}}
     \hline
     Variable & Description & Value & Units \\ [0.5ex] 
     \hline
     $T_{in}$ & Working Fluid Inlet Temperature  & \{-18,18\} & $\degree C$ \\ 
     $T_{air}$ & Air Temperature & 18 & $\degree C$ \\
     $T_{sat}$ & Saturation Temperature & 0 & $\degree C$ \\
     $\dot{m}_{wf}$ & Mass Flow Rate & 0.10 & $kg/s$ \\
     $C_{p,wg}$ & Specific Heat ($wg$) & 3.4 & $kJ/(kg \degree C)$ \\
     $h_{wg}$ & Convective Heat Transfer Coeff. ($wg$) & $10^{4}$ & $ W/( m^2\degree C) $ \\
     $\rho_{wg}$ & Density ($wg$) & 1090 & $kg/m^3$ \\
     $C_{p,inn.}$ & Specific Heat ($Cu$) & 0.39 & $kJ/(kg \degree C)$ \\
     $k_{inn.}$ & Thermal Conductivity ($Cu$) & 401 & $ W/( m \degree C) $ \\
     $\rho_{inn.}$ & Density ($Cu$) & 8960 & $kg/m^3$ \\
     $h_f$ & Heat of Fusion & 334 & $kJ/kg$ \\
     $C_{p,S}, C_{p,L}$ & Specific Heat (Solid, Liquid) & 2.11, 4.18 & $kJ/(kg \degree C)$ \\
     $\rho_S, \rho_L$ & Density (Solid, Liquid) & 916, 1000 & $kg/m^3$ \\
     $k_S, k_L$ & Thermal Conductivity (Solid, Liquid) & 2.3, 0.58 & $W/(m \degree C)$ \\
     $h_{air}$ & Convective Heat Transfer Coeff. ($air$) & $5$ & $ W/( m^2\degree C) $ \\
     $L$ & Length of pipe & 1.00 & $m$ \\
     $R_{ins.}$ & Insulation Resistance & $10^{14}$ & $ \degree C / W $ \\
     $C_{p,out.}$ & Specific Heat ($PVC$) & 0.88 & $kJ/(kg \degree C)$ \\
     $k_{out.}$ & Thermal Conductivity ($PVC$) & 0.20 & $ W/( m \degree C) $ \\
     $\rho_{out.}$ & Density ($PVC$) & 1350 & $kg/m^3$ \\
     $r_1$ & Inner Wall Radius & $6.0$ & $ mm $ \\
     $r_5$ & Outer Wall Radius & $28.8$ & $ mm $ \\
     $\Delta r_{inn.}$ & Inner Wall Thickness & $0.8$ & $ mm $ \\
     $\Delta r_{out.}$ & Outer Wall Thickness & $6.4$ & $ mm $ \\
     $\Delta r_{PCM}$ & PCM Thickness & $19.1$ & $ mm $ \\
     $\Delta r$ & Difference in PCM radii & $0.55$ & $ mm $ \\
     $M_{tot}$ & Total mass of PCM & 1.90 & $kg$ \\ [1ex]
     \hline
    \end{tabular}
    \end{table}
While $ SOC $ is a state of the MB model, for the FG model, $SOC$ is computed as
\begin{equation}
    SOC_{FG} = \frac{1}{M_{tot}}\sum_{j=3}^{n+2} C_j \left(1-\frac{max(0,min(h_j,h_f))}{h_f}\right).
\end{equation}

To compare the SOC for the FG and MB approach, the absolute difference is computed as
\begin{equation}
    \Delta_{SOC} = |SOC_{MB} - SOC_{FG}|.
\end{equation}

\subsection{Complete Freezing and Melting}

Fig. \ref{fig:Freeze_Melt_Plot.pdf} shows the simulation results for the FG and MB models for two complete freezing and melting cycles. The top left plot shows trajectories for the simulated SOC using the FG and MB models, denoted $ SOC_{FG} $ and $ SOC_{MB} $, while the top right plot shows $\Delta_{SOC}$. Notably, the maximum and average values of $ \Delta_{SOC} $ are $ 0.05 $ and $ 0.03 $, with the total computation of $ 1.2 $ seconds for the MB model and $ 6.1 $ seconds for the FG model. The lower left plot in Fig. \ref{fig:Freeze_Melt_Plot.pdf} shows the mode switching for the MB model and the lower right plot shows the solid and liquid region enthalpies for the MB model and all $ n = 35 $ enthalpies of for the PCM of the FG model. While simulating roughly five times faster than the FG model, the MB model is remarkably accurate when simulating complete freezing and melting cycles.


\begin{figure*}[t]
        \centering
        \includegraphics[width=\textwidth]{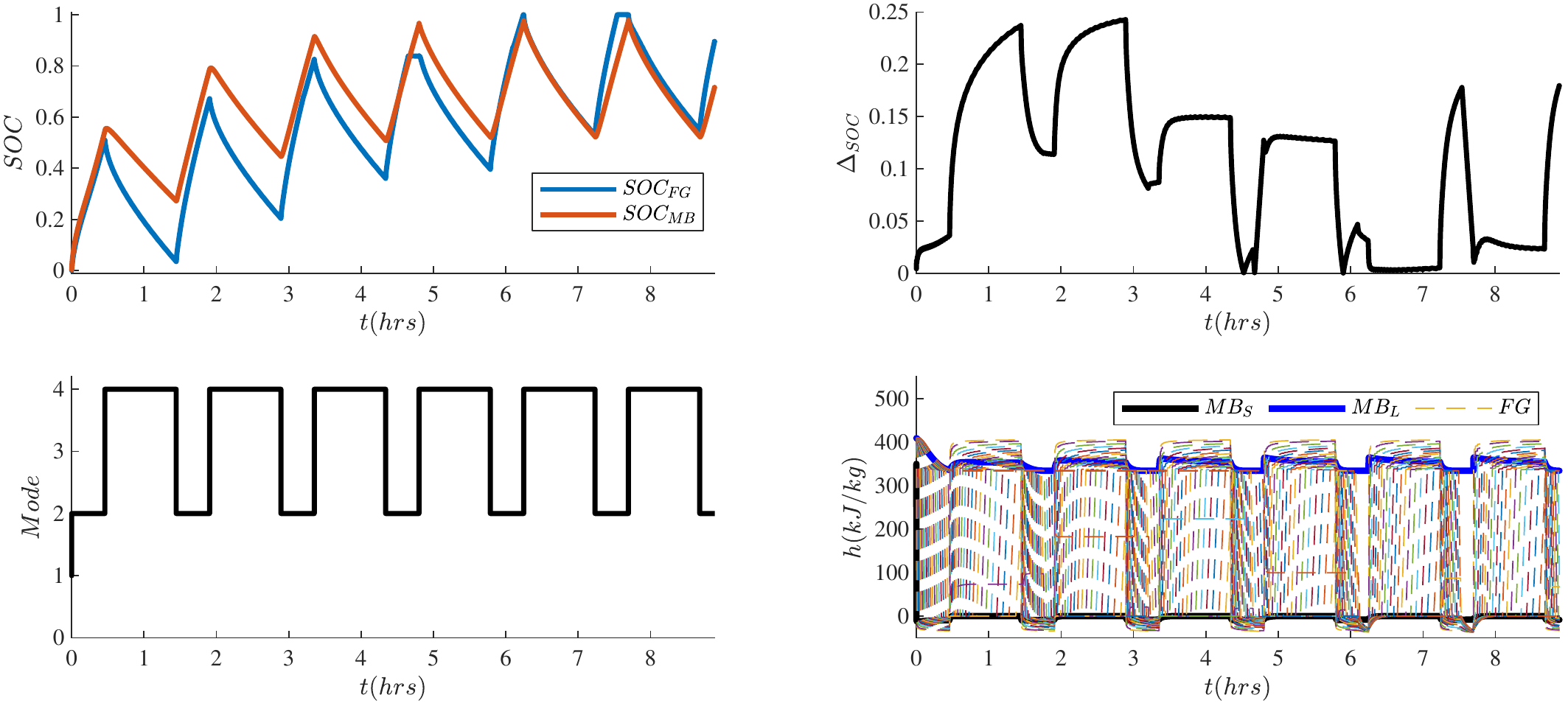}
        \caption{Differences between FG (with $n = 35$) and MB models for six partial freezing and melting cycles.}
        \label{fig:Partial_Freeze_Melt_Plot.pdf}
        \end{figure*}

\subsection{Partial Freezing and Melting}

Fig. \ref{fig:Partial_Freeze_Melt_Plot.pdf} shows how the MB approach loses accuracy when simulating partial freezing and melting of the PCM. When switching between Modes 2 and 4, the maximum value of $ \Delta_{SOC} $ increases significantly up to $ 0.25 $. This is due to the fact that partial freezing can create complex geometries with multiple regions of solid and liquid.  This complex geometry, which results in additional heat transfer surface area between the solid and liquid regions, cannot be captured by the proposed MB model. This is why the $ SOC $ decreases significantly faster using the FG model during the first period of operation in Mode 4.  Note that the accuracy of the MB increases on average when the PCM freezes completely in the latter half of the simulation. In summary, the proposed MB modeling approach is only recommended when complete freezing and melting of the PCM is expected and future work will focus on modifying the MB formulation to more accurately capture behavior associated with partial freezing and melting.

\section{Conclusions}
This paper presented a switched moving boundary approach as a control-oriented method for modeling thermal energy storage devices with phase change material. Graph-based modeling was used to identify the structure of the dynamics when using a fixed grid and the proposed moving bounding modeling approaches. A finite state machine allowed the moving boundary model to switch modes to capture the dynamics associated with freezing, melting, completely solid, and completely liquid Phase change material. A numerical example demonstrated the accuracy and computational efficiency of the switched moving boundary model. 


Future work will focus on experimental validation, capturing the effects of natural convection heat transfer in the liquid regions, and accurately modeling partial freezing and melting operation. Additionally, the switched moving boundary approach will also be extended to model thermal energy storage devices in three dimensions where heat transfer is no longer uniform along the length of the device.

\bibliography{./main_formatted}

\end{document}